\documentclass[3p,times,twocolumn]{elsarticle}
 \biboptions{comma,sort&compress}
 
\usepackage{graphicx}
\usepackage{slashed}
\usepackage{amsmath}
\usepackage{amssymb}
\newcommand{\bs}{\boldsymbol}

\usepackage[colorlinks,citecolor=DarkGreen,linkcolor=DarkRed,urlcolor=DarkBlue]{hyperref}

\usepackage{amssymb}

\usepackage[figuresright]{rotating}

\date{}

\begin{document}

 \begin{frontmatter}

\title{Magnetic and density effects on the nucleon axial coupling }
 \cortext[cor0]{Talk given at 26th International Conference in Quantum Chromodynamics (QCD 23),  10- 14 July  2023, Montpellier - FR}
 \author[label1]{C. A. Dominguez}
\address[label1]{Centre for Theoretical \& Mathematical Physics, and Department of Physics, University of Cape Town, Rondebosch 7700, South Africa.}

 \author[label1,label2,label3]{M. Loewe \fnref{fn1}}
  \fntext[fn1]{Speaker, Corresponding author.}
 \address[label2]{Facultad de Ingenier\'ia, Arquitectura y Diseño, Universidad San Sebasti\'an, Santiago, Chile.}
 \address[label3]{Centro Cient\'ifico Tecnol\'ogico de Valparaíso-CCTVAL,
Universidad T\'ecnica Federico Santa Mar\'ia, Casilla 110-V, Valpara\'iso, Chile.}
\ead{mloewelo@yahoo.com}

 \author[label4]{C. Villavicencio}
  

\address[label4]{Centro de Ciencias Exactas \& Departamento de Ciencias B\'asicas, Facultad de Ciencias, Universidad del B\'io-B\'io,
Casilla 447, Chill\'an, Chile.}
 \author[label5,label6]{R. Zamora}
 \address[label5]{Instituto de Ciencias B\'asicas, Universidad Diego Portales, Casilla 298-V, Santiago, Chile.}
\address[label6]{Centro de Investigaci\'on y Desarrollo en Ciencias Aeroespaciales (CIDCA), Academia Politécnica Aeronáutica, Fuerza A\'erea de Chile, Casilla 8020744, Santiago, Chile.}

\begin{abstract}
 Using appropriate  QCD finite energy sum rules, we discuss the influence of an external magnetic field and baryonic density on the axial-vector coupling constant $g_{A}$. This scenario corresponds to a magnetar environment. We found that $g_{A}$ decreases both as function of the magnetic field strength and the baryonic density. It turns out that at the nuclear density $\rho _{0}$ the axial-vector coupling takes the value $g_{A}^{*}  \approx 0.92$. Although $g_{A}$ decreases in general with the magnetic field intensity, $g_{A}^{*}$ does not change in a relevant way with the magnetic field.  
\end{abstract}



 \end{frontmatter}
\section{Introduction}

The behavior of hadronic matter under extreme conditions has called the attention of people during many years. Initially, the analyses concentrated on two scenarios. In the first case the response of hadronic parameters like masses, widths, effective coupling constants, form factors, etc. to temperature and magnetic fields was discussed. This case corresponds to peripheral heavy ion collisions measured in RHIC and also LHC. In the second approach,  the behavior of hadronic matter is discuss in terms of temperature and baryonic density as the dominant external agents.  Normally, the usual QCD phase diagram is precisely presented in terms of the last parameters.
We expect to have soon measurements related to density effects in NICA and in FAIR. In the present article we will address the situation where baryonic density and magnetic field strength are dominant, concentrating on the evolution, as function of the magnetic field intensity and the baryon density, of the axial-vector coupling $g_{A}(\rho, B)$.  Temperature will not be taken into account in our discussion. This case corresponds to the scenario of magnetars, a certain type of fascinating neutron stars.

\smallskip
This coupling is  relevant for the neutrino emissivity of neutron stars since it is related to the nuclear beta decay processes. The neutron decay width is proportional to $1 + 3g_{A}^2$. \cite{Mund:2012fq,Czarnecki:2018okw}. In fact, we have also  the inverse process, the capture of electrons. Both effects are known altogether as URCA process. \cite{Yakovlev:2000jp}. Different analyses have shown that $g_{A}$ is quenched in nuclear medium. 

\smallskip 
Our analysis will be based on the finite energy sum rules (FESR) approach, involving a three current correlator (proton, axial vector, neutron). This will allow us to determine the magnetic and density evolution of the axial-vector coupling constant. We will need also a set of FESR for two-nucleon correlators (proton-proton and neutron-neutron). We need these correlators to find the dependence of the nucleon-current coupling on the medium properties as well as the evolution of the continuum hadronic threshold.

\section{Finite Energy Sum Rules}

Different types of QCD sum rules have been used in literature as, for example Laplace, Gaussian or Hilbert sum rules. Among them, finite energy QCD sum rules allow to establish a clear criterion to separate the hadronic and QCD sectors. The idea is to integrate a form factor $\Pi(s)$, associated to a certain current correlator, along a circle in the complex s-plane which has a cut on the positive real axis. This has been called the pac-man contour. There are no poles inside the contour. The idea is that hadronic resonances are on the cut whereas the QCD sector is on the circle.The QCD sector includes both perturbative and non-peturbative termswhich expressed as condensates in an operator product expansion (OPE). Cauchy's theorem implies then  immediately

\begin{equation}
\int_0^{s_0}\frac{ds}{\pi} s^N \text{Im}\Pi^\mathrm{\tiny had}(s)
   =-\oint_{s_0}\frac{ds}{2\pi i}s^N\Pi^\mathrm{\tiny QCD}(s) \,,     \end{equation}
\noindent where $s_0$ is the radius of the circle. The analytic kernel $S^N$ does not alters this relation. As it is well known, the OPE is a parameterization of non-perturbative effects in terms of condensates $\langle O_{n}\rangle$ which include the quark and gluon condensates and higher order contributions according to
\begin{equation}
    \Pi(s) = C_\mathrm{\tiny pert}(s) + \sum_{n>0}C_{2n}(s)\frac{\langle O_{2n}\rangle}{s^n},
\end{equation}
In absence of radiative corrections and medium effects, there is a one to one correspondence between the different values of N and the relevant condensates. This correspondence is spoiled in particular by medium effects which break Lorentz symmetry. In these type of calculations it is necessary to choose a frame of reference to perform the integrals since now $\Pi(p^2)$ becomes $\Pi(p_{0},\vec{p})$. Calculations are done, normally, in the frame at rest respect to the vacuum, i.e. $\vec{p} = 0$, taking then $s = p_0 ^2$ and separating the form factor into even and odd components according to $\Pi (p_0) = \Pi ^e(s) + p_{0} \Pi ^o (s)$.

\smallskip
When dealing with the three point  nucleon\,--\,axial-vector\,--\,nucleon, correlator a double FESR appears. \cite{nosotros}. 

\begin{eqnarray}
\int_0^{s_p}\frac{ds'}{\pi} \,\mathrm{Im}_{s'}\!\!\int_0^{s_n}\frac{ds}{\pi}\,\mathrm{Im}_s\Pi^\text{\tiny had} (s,s',t)\nonumber\\
 \; \;\;\;\;=\oint_{s_p}\frac{ds'}{2\pi i}\oint_{s_n}\frac{ds}{2\pi i}\,\Pi^\text{\tiny QCD}(s,s',t),
 \label{eq.FESR_had=QCD}
\end{eqnarray}

where 
\begin{equation}
 \mathrm{Im}_s f(s)\equiv\lim_{\epsilon\to 0}\mathrm{Im}f(s+i\epsilon).
\end{equation}

\smallskip For details, consult the original article \cite{nosotros}. In what follows we will treat first the vacuum case, including then in a second step only density effects taking, finally, the situation where the magnetic field is also consider.

\section{The vacuum sector}

The $g_{A}$ axial-vector coupling will be associated to the currents correlator
\begin{equation}
 \Pi_\mu(x,y,z)=-\langle 0|\,{\cal T}\,\eta_p(x)A_\mu(y)\,\bar\eta_n(z)\,|0\rangle. 
\end{equation}
In the QCD sector $\eta_N$ is Ioffe's nucleon interpolating current \cite{Ioffe:1981kw}

\begin{eqnarray}
 \eta_p(x) &=\epsilon^{abc}\left[u^a(x)^T\,C\gamma^\mu \,u^b(x)\right]\gamma_\mu\gamma_5 \,d^c(x),\\
 \bar\eta_n(z) &= \epsilon^{abc}\left[\bar d^b(z)\,\gamma^\mu C\,\bar d^a(z)^T\right]\bar u^c(z)\,\gamma_\mu\gamma_5 ,
\end{eqnarray}

where $C=i\gamma_0\gamma_2$ is the charge conjugation operator. 

\smallskip In the hadronic sector we introduce phenomenological current nucleon couplings $\lambda_{p}$ and $\lambda  _{n}$, for proton and neutron respectively, according to  
\begin{eqnarray}
 \langle 0|\,\eta_p(x)\,|p',s'\rangle &= \lambda_p\, u_p^{s'}(p')\,e^{-ip'\cdot x},\label{eq.eta_p}\\ 
 \langle p,s|\,\bar\eta_n(z)\, |0\rangle &= \lambda_n\, \bar u_n^s(p)\, e^{ip\cdot z}.
\end{eqnarray}
In the QCD sector, the axial-vector current operators is given by
\begin{equation}
A_\mu(y) = \bar d(y)\,\gamma_\mu\gamma_5\, u(y).
\end{equation}

Going oing now into the hadronic sector, the matrix element of the axial-vector current in momentum space can be expressed as
\begin{equation}
 \langle p',s'|A_\mu(y)|p,s\rangle = \bar u_p^{s'}(p')\,T_\mu(q)\,u_n^s(p) \,e^{iq\cdot y},
\end{equation}
with $q=p'-p$.

\smallskip By expanding the function T in the above equation in the Clifford basis, in the most general way compatible with the quantum numbers in the axial-vector sector, we find 
\begin{eqnarray}
 T_\mu(q) = 
 G_A(t)\gamma_\mu\gamma_5 +G_P(t)\gamma_5 \frac{q_\mu}{2  m_N}\\
 \nonumber
 +G_T(t)\sigma_{\mu\nu}\gamma_5  \frac{q_\nu}{2 m_N},
 \label{T_mu}
\end{eqnarray}
 with $t=q^2$  and  $ m_N$ the vacuum nucleon mass. The axial vector coupling corresponds to $g_{A} \equiv G_{A}(0)$.

 \smallskip It turns out that the correlator $\Pi _{\mu}$ in momentum space, once a complete set of intermediate nucleon states is inserted, can be written as 

\begin{equation}
 \Pi^\text{\tiny had}_\mu(p,p')=\lambda_n\lambda_p\frac{ (\slashed{p}+m_n)T_\mu(q)(\slashed{p}'+m_p)}{(p^2-m_n^2)(p'^2-m_p^2)}.
\end{equation}
It is possible to isolate $G_{A}$ form the other terms in the general decomposition  of $T_{\mu}$ using
\begin{equation}
 \mathrm{tr}\,[\Pi_\mu(p,p')\,\gamma_\nu]=-4i\epsilon_{\mu\nu\alpha\beta}p^\alpha p'^\beta \Pi(s,s',t),
\end{equation}
with $s=p^2$, $s'=p'^2$. $\Pi$ in the hadronic sector is given by 
\begin{equation}
 \Pi^\text{\tiny had}(s,s',t)=\lambda_n\lambda_p\frac{G_A(t)+G_T(t)(m_n-m_p)/ m_N}{(s-m_n^2)(s'-m_p^2)}.
 \label{Pi_had}
\end{equation}

Since the masses of proton and neutron are almost identical, this is related to the $SU(2)$ isospin symmetry, the contribution from $G_{T}$ can be neglected.
We notice, however, that if isospin density effects are considered,  the mass difference between proton and neutron could be important.

\smallskip
Following the philosophy of QCD sum rules, we have to repeat the same procedure in the QCD sector. By taking the results found  in \cite{Villavicencio:2022gbr} for the two-loop leading order perturbative contribution when $t=0$ we find

\begin{eqnarray}
 \Pi^\text{\tiny pQCD}(s,s',0)= 
 \frac{s^2\ln(-s/\mu^2)-s'^2\ln(-s'/\mu^2)}{(2\pi)^4\,(s'-s)}\\
 +\text{regular \;\; terms}\nonumber.
 \label{PI-pQCD}
\end{eqnarray}

 In the above expression $\mu$ is the $\overline{\text{MS}}$ scale. Regular terms are eventual contributions without discontinuities on the real axes, or singularities which would vanish in a FESR framework.

 \smallskip
 Using the double FESR mentioned before we find
 \begin{eqnarray}
 g_A\lambda_n\lambda_p
=\frac{1}{48\pi^4}\left[s_n^3\,\theta(s_p-s_n)
+s_p^3\,\theta(s_n-s_p)\right]
\label{eq.gA(p,n)}
\end{eqnarray}
where $s_N$ are the nucleon current hadronic thresholds, which must be bigger than $m_N^2$. In vacuum, since the proton and neutron thresholds are equal we have
\begin{equation}
g_A=\frac{s_0^3}{48\pi^4\lambda_N^2}.
\end{equation}
This is the leading contribution. The next correction, proportional to the gluon condensates, is negligible.

\smallskip
The two point nucleon-nucleon correlator 
\cite{Ioffe:1981kw,Reinders:1984sr,Sadovnikova:2005ye,Nasrallah:2013ywh}
\begin{equation}
\Pi_N(x)=\langle 0|\,{\cal T}\,\eta_N(x)\,\bar\eta_N(0)\,|0\rangle.
\end{equation}
provides us information about the $\lambda$ couplings. From the corresponding FESR we find 
\begin{equation}
\lambda_N^2 = \frac{s_0^3}{192\pi^4}+\frac{s_0}{32\pi^2}\langle G^2\rangle+\frac{2}{3}\langle\bar qq\rangle^2
\end{equation}
\label{eq.Nuclear_FESR_p}
and
 \begin{equation}\lambda_N^2m_N =-\frac{s_0^2}{8\pi^2}\langle\bar qq\rangle+ \frac{1}{12}\langle G^2\rangle\langle\bar qq\rangle,
 \label{eq.Nuclear_FESR_s}
\end{equation}
where the saturation approximation was invoked in operators of dimension d=6 and d=7. As was pointed out in \cite{Villavicencio:2022gbr}, there is a small window for possible values of quark and gluon condensates  agreeing with the experimental values of the axial-vector coupling constant $g_A\approx 1.275$. 

\smallskip In the presence of medium effects we expect the appearance of several components in the contribution of the axial-vector to the form factor. In our case, density and magnetic corrections will be associated to the in-medium nucleon-nucleon current couplings as well as to the nucleon continuum hadronic thresholds.

\section{Finite Density Effects}

In literature we find a series of papers dealing with finite bearyonic density effects in the baryonic sector \cite{Cohen:1991js}. Density effects will affect effective operators in the non-perturbative sector. New condensates appear. It turns out that all  correlator structures can be separated into  {\it even}  and  {\it odd}   contributions in terms of $p_0$ according to

\begin{align}
\Pi(p_0,\bs{p}^2) &= 
\Pi^e(p_0^2,\bs{p}^2)+p_0\Pi^o(p_0^2,\bs{p}^2).
\label{eq.even-odd}
\end{align}

Introducing of the four velocity $u^{\mu}$, it reduces to $u ^{\mu} = (1,0,0,0)$ in the system's rest frame, the nucleon-nucleon correlator can be decomposed as
\begin{equation}
    \Pi_N = \Pi_s+\slashed{p}\Pi_p+\slashed{u}\Pi_u \,
\end{equation}
where the subscript $s$ refers to the scalar part and the subscript $p$ means proportional to $p$. 

\smallskip In the nuclenoic sector we might parameterize the nucleon correlator as
\begin{equation}
    \Pi_N = \frac{-\lambda_N^2}{\gamma_0 (p_0 +\Delta\mu)-v\,\bs{\gamma}\cdot \bs{p} -m_N},
    \label{eq.Pi_N-had}
\end{equation}
where we have introduced $\Delta \mu $ as a correction to the baryon chemical potential beeing $v$ is the fermion velocity.
Neglecting isospin breaking terms, proton and neutron correlators will lead to the same results.For the present analysis we should concentrate only on two of all possible terms in order to find $\lambda _{N}$ avoiding unknown higher dimensional  condensates. Using Ioffe's version for the interpolating nucleon current  we find, up to dimension 6
\cite{Cohen:1991js,Cohen:1991nk,Furnstahl:1992pi,Jin:1993up,Jeong:2012pa,Ohtani:2016pyk,Cai:2019vsg}

\begin{align}
\Pi_p^e &=    -\frac{1}{64\pi^4}s^2\ln(-s)
    + \frac{1}{9\pi^2}(4\langle \theta_u\rangle + \langle\theta_d\rangle)
    \nonumber\\ & \qquad
    -\frac{1}{32\pi^2}\ln(-s)\langle G^2\rangle+
    \frac{1}{36\pi^2}\ln(-s)\langle \theta_g\rangle
    \nonumber\\&\qquad
    -\frac{2}{3}\langle\bar uu\rangle^2
    -\frac{4}{3}\langle u^\dag u\rangle^2
     +r.t. \label{eq.Pi^e(rhoB)}
     \end{align}
    and
\begin{eqnarray}
\Pi_s^e &=\frac{1}{4\pi^2}\ln(-s)\langle \bar dd\rangle 
    -\frac{1}{12}\langle G^2\rangle\langle\bar dd\rangle
    +r.t..
    \end{eqnarray}
In the previous equations $\langle\theta _{q}\rangle$ is the zero component of the quarks energy momentum tensor and $\langle \theta _{g}\rangle$ refers to the trace of the gluon energy momentum tensor. 

We have taken the following well established density dependent values 

\begin{align}
 \langle q^\dag q\rangle &= \frac{3}{2}\rho_B\\
  \langle\bar qq\rangle &= \langle\bar qq\rangle_0 [1-0.329(\rho_B/\rho_0)] \\
\langle G^2\rangle &=\langle G^2\rangle_0[1-0.066(\rho_B/\rho_0)]\\
\langle \Theta_q\rangle &= 2.6\times 10^{-4}(\rho_B/\rho_0)\\
\langle \Theta_g\rangle &=  6.1\times 10^{-5}(\rho_B/\rho_0)\\
  m_N^* &= m_N[1-0.329(\rho_B/\rho_0)]
  \end{align}
where $\rho_B$ is the baryon density, $\rho_0=0.16 \mathrm{\, fm}^{-3} = 1.22\times 10^{-3}\mathrm{\, GeV}^3$ is the nuclear density.
In the estimations considered above, we have used the value of the pion-nucleon sigma term as $\sigma_N=45\textrm{ MeV}$

\section {Magnetic Field Effects}
 Propagators are corrected by magnetic field effects and in principle also the condensates. Here we have used the weak field expansion for fermionic propagators. When density and magnetic field effects are included, we have to deal with many new condensates due to Lorentz symmetry breaking. However, we have restricted the condensates to the relevant one at finite density. In fact, the only magnetized condensate we have considered  is the polarization of the chiral condensate $\langle \bar q\sigma_{12} q\rangle$. Besides the chiral condensate there is not much known in literature about condensates when density and magnetic field corrections are considered. For low values of the magnetic field strength, the evolution of the gluon condensate is negligible \cite{DElia:2015eey,Dominguez:2018njv}.
In this way, it is reasonable to consider only a baryon density dependent gluon condensate. 
From \cite{Hutauruk:2021dgv} we can see that for low $B$, the approximation $f(B,\rho_B)\approx f(B,0)f(0,\rho_B)/f(0,0)$ can be applied to the quark condensate. 
We will use this approximation also for the nucleon mass, with the magnetic evolution of chiral condensate and nucleon mass the same used in \cite{Dominguez:2020sdf}.
In the case of  $\langle \theta_q\rangle $ and $\langle \theta_g\rangle $, we will left them as baryon density dependent only.

 \smallskip
 Our set of FESR for the proton are given by
 \begin{align}
 \lambda_p^2 &=  \frac{s_p^3}{192\pi^4}  
    + \frac{s_p}{32\pi^3}\langle\alpha_s G^2\rangle 
    + \frac{2}{3}\langle\bar uu\rangle^2 
    \nonumber\\&\quad
        +\frac{s_p}{2\pi^4}e_u e_d B^2
    +\frac{ s_p}{6\pi^4}(e_uB)^2[\ln(s_p/8m_q^2)-1]
   \nonumber\\&\quad  
    +\frac{s_p}{96\pi^4}(e_dB)^2\left[8\ln(s_p/8m_q^2)-9\right]
    \nonumber\\&\quad
    +3\rho_B^2
  - \frac{s_p}{9\pi^2} \langle \theta_d\rangle
  -\frac{4s_p}{9\pi^2}\langle \theta_u\rangle
  -\frac{s_p}{72\pi^2}\langle \theta_g\rangle
  \label{eq.lambda_p-B,rho1,}
  \end{align}

  \begin{eqnarray}
  \lambda_p^2 m_p  &= -\frac{s_p^2}{8\pi^2}\langle\bar dd\rangle
    +\frac{1}{12\pi}\langle \alpha_sG^2\rangle\langle\bar dd\rangle
        \nonumber\\&\quad
    +\frac{s_p}{2\pi^2}e_uB\langle \bar d\sigma_{12}d\rangle
    \nonumber\\&\quad
    +\frac{4}{3\pi^2}(e_u B)^2 \left[\ln(s_p/m_q^2)-1\right]\langle\bar dd\rangle,
    \label{eq.lambda_p-B,rho2}
    \end{eqnarray}
    and
  \begin{eqnarray}
  -\lambda_p^2 \frac{\kappa_p B}{2} &= \frac{s_p^2}{48\pi^2}\langle \bar d\sigma_{12}d\rangle
   + \frac{e_u B\,s_p}{24\pi^2}\langle\bar dd\rangle \label{eq.lambda_p-B,rho3}
  \end{eqnarray}

  \smallskip We have the same set of FESR for the neutron case with the replacement $p \rightarrow n$ and $u \leftrightarrow d$. In the above equations, $e_{u}$, and $e_{d}$ are the quark electric charges, $s_{p}$ and $s_{n}$ ar the continuum proton and neutron thresholds, respectively and $\kappa_{p}$ is the proton anomalous magnetic moment. Our results show that the $\lambda $ couplings decrease with the baryonic density and increase with the magnetic field strength. Going into the continuum thresholds, both $s_{p}$ and $s_{n}$ diminish as function of the baryonic density whereas they increase with the magnetic field. In this sense the magnetic field gas a confining effect. Quite interesting is the fact that always $s_{n} < s_{p}$. We ask the reader to consult the original article for details. This means that the axial-vector coupling constant at finite baryon density and magnetic field is given by

  \begin{equation}
    g_A=\frac{1}{48\pi^4}\frac{s_n^3}{\lambda_n\lambda_p}.
\end{equation}
Since we have, independently, obtained the medium evolution of the couplings $\lambda _{p}$ and $\lambda _{n}$ and the continuum threshold $s_{n}$, being this always smaller than the proton threshold, the previous equation provides us with the evolution of the axial-vector coupling constant. In Fig. 1 we present the decreasing evolution of $g_{A}$ with the baryon density normalized by the nuclear density for three different values of the magnetic field. Quite interesting is the fact that its value at the nuclear density $g_A^*\equiv g_A(\rho_0)   \approx 0.92$ in agreement with other approaches that suggest that $g_A^* \lesssim 1$. In Fig.2 we show the behavior of the axial-vector coupling with the magnetic field strength. As the magnetic field increases, the axial-vector coupling becomes smaller for $\rho_B\lesssim 1.25\rho_0$. 
However, for $\rho \gtrsim 1.25 \rho_0$, $g_A$ seems to increase with the magnetic field.
This means in magnetar's nuclear matter, $g_A \sim g_A^*$  is apparently not affected by the magnetic field which should increase when going to the center of the star.

  \begin{figure}
\includegraphics[scale=0.40]{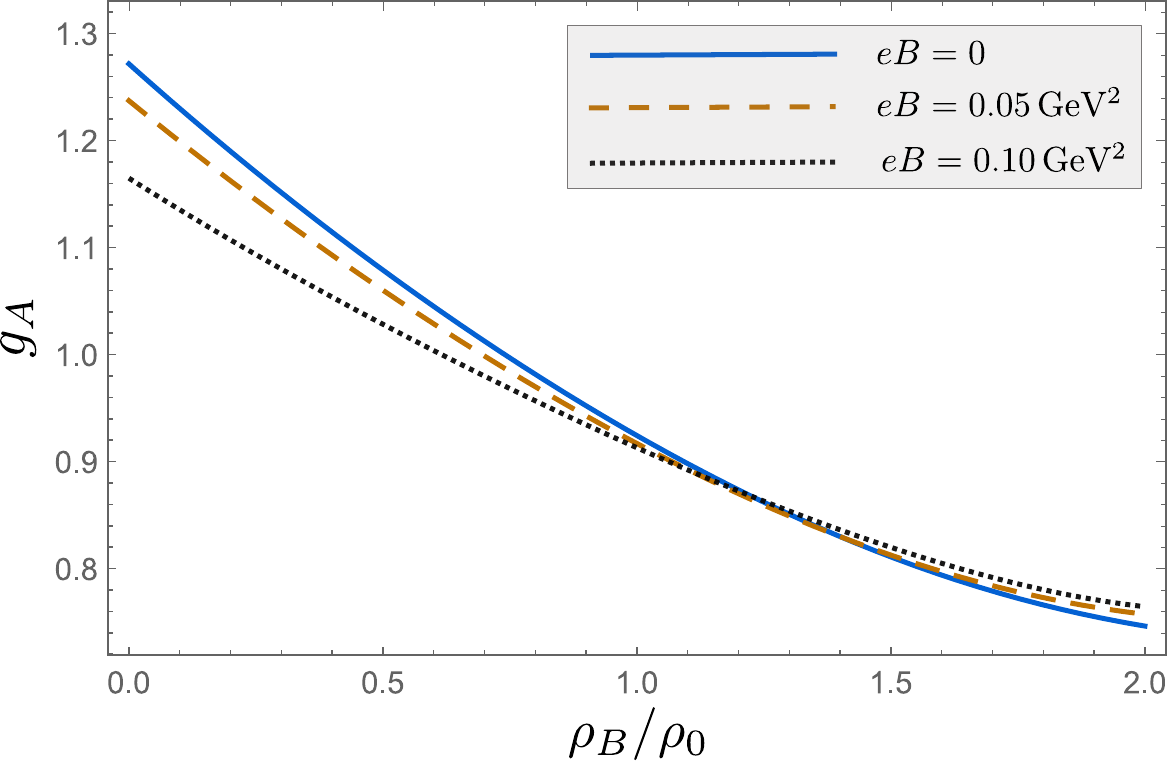}
\caption{Nuclear axial-vector coupling constant as function of the baryon density in units of the nuclear density for different values of the external magnetic field strength.}
\label{fig.gA_rho}
\end{figure}

\begin{figure}
\includegraphics[scale=0.47]{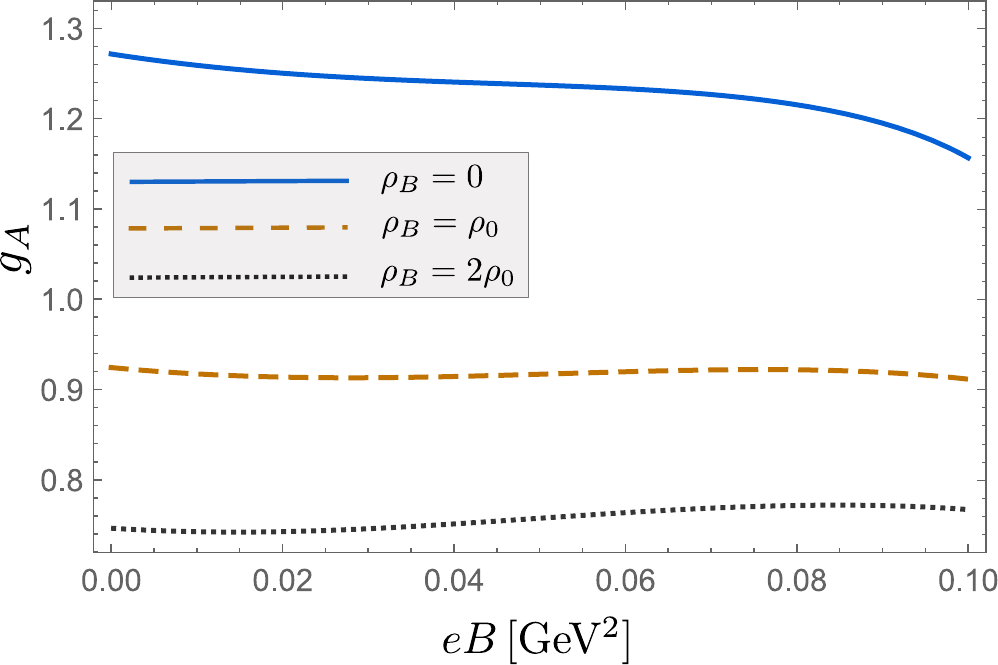}
\caption{Nuclear axial-vector coupling constant as function of the the external magnetic field strength for different values of baryon density in units of the nuclear density.}
\label{fig.gA_eB}
\end{figure}

\section{Conclusions}
In the frame of appropriate FESR we have presented a detailed discussion of the medium evolution, i.e. as function of baryonic density and magnetic field strength, of the axial-vector coupling constant. Our work aims to emulate the scenario of nuclear matter in magnetars. For this purpose we need the corresponding evolution of the nucleonic couplings $\lambda$ and of the continuum thresholds for proton and neutrons. The determination of the axial-vector coupling
$g_{A}$ is related to a three point correlator.  

\smallskip When the baryonic density increases the axial-vector coupling diminishes. In particular, for the nuclear density $\rho _{B}$ we found $g_A^* \approx 0.92$ in accordance with the commonly assumed value that establishes that $g_A^*\sim 1$ \cite{Rho:1974cx,Wilkinson:1974huj,Brown:1978zz,Park:1997vv,Carter:2001kw,Lu:2001mf,Suhonen:2017krv,Bass:2020bkl,Rho:2021zwm,Rho:2023vow}.

The last value seems to be quite stable respect to variations of the magnetic field. It is interesting to stress that the different plotted curves shown in Fig.1 have an intersection. The increasing behavior of $g_{A}$ with the magnetic field should be enhanced for higher values of the baryonic density.

\smallskip As a final remark, we notice that there are many approximations involved in the magnetic evolution of the different condensates. In a future work we would like to obtain complete and better determination of these condensates ant to explore also the role of other condensates neglected in our discussion. 

\section{Acknowledgements}

M.L., C.V. and R.Z. acknowledge support from ANID/CONICYT FONDECYT Regular (Chile) under Grants No. 1190192, 1200483 and 1220035. M. L. acknowledges also support from ANID PIA/APOYO AFB 180002 (Chile) and from the Programa de Financiamiento Basal FB 210008 para Centros Cient\'{\i}ficos y Tecnol\'ogicos de excelencia de ANID


\begin{thebibliography}{10}
\expandafter\ifx\csname url\endcsname\relax
  \def\url#1{\texttt{#1}}\fi
\expandafter\ifx\csname urlprefix\endcsname\relax\def\urlprefix{URL }\fi
\expandafter\ifx\csname href\endcsname\relax
  \def\href#1#2{#2} \def\path#1{#1}\fi

\bibitem{Mund:2012fq}
D.~Mund, B.~Maerkisch, M.~Deissenroth, J.~Krempel, M.~Schumann, H.~Abele,
  A.~Petoukhov, T.~Soldner, {Determination of the Weak Axial Vector Coupling
  from a Measurement of the Beta-Asymmetry Parameter A in Neutron Beta Decay},
  Phys. Rev. Lett. 110 (2013) 172502.
\newblock \href {http://arxiv.org/abs/1204.0013} {\path{arXiv:1204.0013}},
  \href {http://dx.doi.org/10.1103/PhysRevLett.110.172502}
  {\path{doi:10.1103/PhysRevLett.110.172502}}.

\bibitem{Czarnecki:2018okw}
A.~Czarnecki, W.~J. Marciano, A.~Sirlin, {Neutron Lifetime and Axial Coupling
  Connection}, Phys. Rev. Lett. 120~(20) (2018) 202002.
\newblock \href {http://arxiv.org/abs/1802.01804} {\path{arXiv:1802.01804}},
  \href {http://dx.doi.org/10.1103/PhysRevLett.120.202002}
  {\path{doi:10.1103/PhysRevLett.120.202002}}.

\bibitem{Yakovlev:2000jp}
D.~G. Yakovlev, A.~D. Kaminker, O.~Y. Gnedin, P.~Haensel, {Neutrino emission
  from neutron stars}, Phys. Rept. 354 (2001) 1.
\newblock \href {http://arxiv.org/abs/astro-ph/0012122}
  {\path{arXiv:astro-ph/0012122}}, \href
  {http://dx.doi.org/10.1016/S0370-1573(00)00131-9}
  {\path{doi:10.1016/S0370-1573(00)00131-9}}.

\bibitem{nosotros}
C.~A. Dominguez, M.~Loewe, C.~Villavicencio, R.~Zamora, {Nucleon axial-vector
  coupling constant in magnetar environments}\href
  {http://arxiv.org/abs/2308.05663} {\path{arXiv:2308.05663}}.

\bibitem{Ioffe:1981kw}
B.~L. Ioffe, {Calculation of Baryon Masses in Quantum Chromodynamics}, Nucl.
  Phys. B 188 (1981) 317--341, [Erratum: Nucl.Phys.B 191, 591--592 (1981)].
\newblock \href {http://dx.doi.org/10.1016/0550-3213(81)90259-5}
  {\path{doi:10.1016/0550-3213(81)90259-5}}.

\bibitem{Villavicencio:2022gbr}
C.~Villavicencio, {Axial coupling constant in a magnetic background}, Phys.
  Rev. D 107~(7) (2023) 076009.
\newblock \href {http://arxiv.org/abs/2212.04649} {\path{arXiv:2212.04649}},
  \href {http://dx.doi.org/10.1103/PhysRevD.107.076009}
  {\path{doi:10.1103/PhysRevD.107.076009}}.

\bibitem{Reinders:1984sr}
L.~Reinders, H.~Rubinstein, S.~Yazaki, {Hadron Properties from QCD Sum Rules},
  Phys. Rept. 127 (1985) 1.
\newblock \href {http://dx.doi.org/10.1016/0370-1573(85)90065-1}
  {\path{doi:10.1016/0370-1573(85)90065-1}}.

\bibitem{Sadovnikova:2005ye}
V.~Sadovnikova, E.~Drukarev, M.~Ryskin, {Nucleon QCD sum rules with the
  radiative corrections}, Phys. Rev. D 72 (2005) 114015.
\newblock \href {http://arxiv.org/abs/hep-ph/0508240}
  {\path{arXiv:hep-ph/0508240}}, \href
  {http://dx.doi.org/10.1103/PhysRevD.72.114015}
  {\path{doi:10.1103/PhysRevD.72.114015}}.

\bibitem{Nasrallah:2013ywh}
N.~F. Nasrallah, K.~Schilcher, {New sum rule determination of the nucleon mass
  and strangeness content}, Phys. Rev. C 89~(4) (2014) 045202, [Addendum:
  Phys.Rev.C 89, 059904 (2014)].
\newblock \href {http://arxiv.org/abs/1310.6114} {\path{arXiv:1310.6114}},
  \href {http://dx.doi.org/10.1103/PhysRevC.89.045202}
  {\path{doi:10.1103/PhysRevC.89.045202}}.

\bibitem{Cohen:1991js}
T.~D. Cohen, R.~J. Furnstahl, D.~K. Griegel, {From QCD sum rules to
  relativistic nuclear physics}, Phys. Rev. Lett. 67 (1991) 961--964.
\newblock \href {http://dx.doi.org/10.1103/PhysRevLett.67.961}
  {\path{doi:10.1103/PhysRevLett.67.961}}.

\bibitem{Cohen:1991nk}
T.~D. Cohen, R.~J. Furnstahl, D.~K. Griegel, {Quark and gluon condensates in
  nuclear matter}, Phys. Rev. C 45 (1992) 1881--1893.
\newblock \href {http://dx.doi.org/10.1103/PhysRevC.45.1881}
  {\path{doi:10.1103/PhysRevC.45.1881}}.

\bibitem{Furnstahl:1992pi}
R.~J. Furnstahl, D.~K. Griegel, T.~D. Cohen, {QCD sum rules for nucleons in
  nuclear matter}, Phys. Rev. C 46 (1992) 1507--1527.
\newblock \href {http://dx.doi.org/10.1103/PhysRevC.46.1507}
  {\path{doi:10.1103/PhysRevC.46.1507}}.

\bibitem{Jin:1993up}
X.-m. Jin, M.~Nielsen, T.~D. Cohen, R.~J. Furnstahl, D.~K. Griegel, {QCD Sum
  rules for nucleons in nuclear matter. 3.}, Phys. Rev. C 49 (1994) 464--477.
\newblock \href {http://dx.doi.org/10.1103/PhysRevC.49.464}
  {\path{doi:10.1103/PhysRevC.49.464}}.

\bibitem{Jeong:2012pa}
K.~S. Jeong, S.~H. Lee, {Nuclear Symmetry Energy from QCD sum rules}, Phys.
  Rev. C 87~(1) (2013) 015204.
\newblock \href {http://arxiv.org/abs/1209.0080} {\path{arXiv:1209.0080}},
  \href {http://dx.doi.org/10.1103/PhysRevC.87.015204}
  {\path{doi:10.1103/PhysRevC.87.015204}}.

\bibitem{Ohtani:2016pyk}
K.~Ohtani, P.~Gubler, M.~Oka, {Negative-parity nucleon excited state in nuclear
  matter}, Phys. Rev. C 94~(4) (2016) 045203.
\newblock \href {http://arxiv.org/abs/1606.09434} {\path{arXiv:1606.09434}},
  \href {http://dx.doi.org/10.1103/PhysRevC.94.045203}
  {\path{doi:10.1103/PhysRevC.94.045203}}.

\bibitem{Cai:2019vsg}
B.-J. Cai, L.-W. Chen, {Relativistic self-energy decomposition of nuclear
  symmetry energy and equation of state of neutron matter within QCD sum
  rules}, Phys. Rev. C 100~(2) (2019) 024303.
\newblock \href {http://arxiv.org/abs/1903.10430} {\path{arXiv:1903.10430}},
  \href {http://dx.doi.org/10.1103/PhysRevC.100.024303}
  {\path{doi:10.1103/PhysRevC.100.024303}}.

\bibitem{DElia:2015eey}
M.~D'Elia, E.~Meggiolaro, M.~Mesiti, F.~Negro, {Gauge-invariant field-strength
  correlators for QCD in a magnetic background}, Phys. Rev. D 93 (2016) 054017.
\newblock \href {http://arxiv.org/abs/1510.07012} {\path{arXiv:1510.07012}},
  \href {http://dx.doi.org/10.1103/PhysRevD.93.054017}
  {\path{doi:10.1103/PhysRevD.93.054017}}.

\bibitem{Dominguez:2018njv}
C.~A. Dominguez, M.~Loewe, C.~Villavicencio, {QCD determination of the magnetic
  field dependence of QCD and hadronic parameters}, Phys. Rev. D 98~(3) (2018)
  034015.
\newblock \href {http://arxiv.org/abs/1806.10088} {\path{arXiv:1806.10088}},
  \href {http://dx.doi.org/10.1103/PhysRevD.98.034015}
  {\path{doi:10.1103/PhysRevD.98.034015}}.

\bibitem{Hutauruk:2021dgv}
P.~T.~P. Hutauruk, S.-i. Nam, {QCD chiral condensate and pseudoscalar-meson
  properties in the nuclear medium at finite temperature}, Mod. Phys. Lett. A
  37~(14) (2022) 2250087.
\newblock \href {http://arxiv.org/abs/2103.09426} {\path{arXiv:2103.09426}},
  \href {http://dx.doi.org/10.1142/S0217732322500870}
  {\path{doi:10.1142/S0217732322500870}}.

\bibitem{Dominguez:2020sdf}
C.~A. Dominguez, L.~A. Hern\'andez, M.~Loewe, C.~Villavicencio, R.~Zamora,
  {Magnetic field dependence of nucleon parameters from QCD sum rules}, Phys.
  Rev. D 102~(9) (2020) 094007.
\newblock \href {http://arxiv.org/abs/2008.10742} {\path{arXiv:2008.10742}},
  \href {http://dx.doi.org/10.1103/PhysRevD.102.094007}
  {\path{doi:10.1103/PhysRevD.102.094007}}.

\bibitem{Rho:1974cx}
M.~Rho, {Quenching of axial-vector coupling constant in beta-decay and
  pion-nucleus optical potential}, Nucl. Phys. A 231 (1974) 493--503.
\newblock \href {http://dx.doi.org/10.1016/0375-9474(74)90512-0}
  {\path{doi:10.1016/0375-9474(74)90512-0}}.

\bibitem{Wilkinson:1974huj}
D.~H. Wilkinson, {Renormalization of the axial-vector coupling constant in
  nuclear \ensuremath{\beta}-decay (III)}, Nucl. Phys. A 225 (1974) 365--381.
\newblock \href {http://dx.doi.org/10.1016/0375-9474(74)90347-9}
  {\path{doi:10.1016/0375-9474(74)90347-9}}.

\bibitem{Brown:1978zz}
B.~A. Brown, W.~Chung, B.~H. Wildenthal, {Empirical Renormalization of the
  One-Body Gamow-Teller beta-Decay Matrix Elements in the 1s-0d Shell}, Phys.
  Rev. Lett. 40 (1978) 1631--1635.
\newblock \href {http://dx.doi.org/10.1103/PhysRevLett.40.1631}
  {\path{doi:10.1103/PhysRevLett.40.1631}}.

\bibitem{Park:1997vv}
T.-S. Park, H.~Jung, D.-P. Min, {In-medium effective axial - vector coupling
  constant}, Phys. Lett. B 409 (1997) 26--32.
\newblock \href {http://arxiv.org/abs/nucl-th/9704033}
  {\path{arXiv:nucl-th/9704033}}, \href
  {http://dx.doi.org/10.1016/S0370-2693(97)00880-0}
  {\path{doi:10.1016/S0370-2693(97)00880-0}}.

\bibitem{Carter:2001kw}
G.~W. Carter, M.~Prakash, {The Quenching of the axial coupling in nuclear and
  neutron star matter}, Phys. Lett. B 525 (2002) 249--254.
\newblock \href {http://arxiv.org/abs/nucl-th/0106029}
  {\path{arXiv:nucl-th/0106029}}, \href
  {http://dx.doi.org/10.1016/S0370-2693(01)01452-6}
  {\path{doi:10.1016/S0370-2693(01)01452-6}}.

\bibitem{Lu:2001mf}
D.-H. Lu, A.~W. Thomas, K.~Tsushima, {Medium modification of the nucleon axial
  form-factor}\href {http://arxiv.org/abs/nucl-th/0112001}
  {\path{arXiv:nucl-th/0112001}}.

\bibitem{Suhonen:2017krv}
J.~T. Suhonen, {Value of the Axial-Vector Coupling Strength in
  \ensuremath{\beta} and \ensuremath{\beta}\ensuremath{\beta} Decays: A
  Review}, Front. in Phys. 5 (2017) 55.
\newblock \href {http://arxiv.org/abs/1712.01565} {\path{arXiv:1712.01565}},
  \href {http://dx.doi.org/10.3389/fphy.2017.00055}
  {\path{doi:10.3389/fphy.2017.00055}}.

\bibitem{Bass:2020bkl}
S.~D. Bass, {Gamow\textendash{}Teller Transitions and the Spin EMC Effect: the
  Bjorken Sum-rule in Medium}, Acta Phys. Polon. B 52~(1) (2021) 43--52.
\newblock \href {http://arxiv.org/abs/2006.10601} {\path{arXiv:2006.10601}},
  \href {http://dx.doi.org/10.5506/APhysPolB.52.43}
  {\path{doi:10.5506/APhysPolB.52.43}}.

\bibitem{Rho:2021zwm}
M.~Rho, Y.-L. Ma, {Manifestation of Hidden Symmetries in Baryonic Matter: From
  Finite Nuclei to Neutron Stars}, Mod. Phys. Lett. A 36~(13) (2021) 2130012.
\newblock \href {http://arxiv.org/abs/2101.07121} {\path{arXiv:2101.07121}},
  \href {http://dx.doi.org/10.1142/S0217732321300123}
  {\path{doi:10.1142/S0217732321300123}}.

\bibitem{Rho:2023vow}
M.~Rho, {Anomaly-Induced Quenching of ${g_A}$ in Nuclear Matter and Impact on
  Search for Neutrinoless $\beta\beta$ Decay}\href
  {http://arxiv.org/abs/2302.00554} {\path{arXiv:2302.00554}}.

\end{thebibliography}

\end{document}